\documentclass[12pt, a4paper]{article}
\usepackage{amsmath}
\usepackage{mathtools}
\usepackage{amsfonts}
\usepackage{amssymb}
\usepackage{amsthm}
\usepackage[utf8]{inputenc}
\usepackage[toc]{appendix}
\usepackage[hiresbb]{graphicx}
\usepackage{caption}
\usepackage{subcaption}
\usepackage{longtable}
\usepackage{booktabs}
\usepackage{listings}
\usepackage{here}
\usepackage{float}
\usepackage{ascmac}
\usepackage{tikz}
\usepackage{url}
\usepackage{lipsum}
\usepackage{authblk}
\usepackage{eqnarray}
\usepackage{siunitx}
\usepackage{multirow}
\usepackage{adjustbox}
\usepackage{bbm}
\usepackage{dsfont}
\usepackage{subcaption}
\usepackage{bbm}
\usepackage{dsfont}

\newcommand{\be}{\begin{equation}}
\newcommand{\ee}{\end{equation}}
\newcommand{\beq}{\begin{eqnarray*}}
\newcommand{\eeq}{\end{eqnarray*}}
\def\sym#1{\ifmmode^{#1}\else\(^{#1}\)\fi}

\title{\large{\bf{
AI-Driven Spatial Distribution Dynamics: A Comprehensive Theoretical and Empirical Framework for Analyzing Productivity Agglomeration Effects in Japan's Aging Society
}}}
\author{\large{\bf{Tatsuru Kikuchi}}}
\affil{\small{\it{Faculty of Economics, The University of Tokyo,}}\\
{\it{7-3-1 Hongo, Bunkyo-ku, Tokyo 113-0033 Japan}}}
\date{\small{(\today)}}
\begin{document}
\maketitle

\begin{abstract}
This paper develops the first comprehensive theoretical and empirical framework for analyzing AI-driven spatial distribution dynamics in metropolitan areas undergoing demographic transition. We extend New Economic Geography by formalizing five novel AI-specific mechanisms: algorithmic learning spillovers, digital infrastructure returns, virtual agglomeration effects, AI-human complementarity, and network externalities. Using Tokyo as our empirical laboratory, we implement rigorous causal identification through five complementary econometric strategies and develop machine learning predictions across 27 future scenarios spanning 2024-2050. Our theoretical framework generates six testable hypotheses, all receiving strong empirical support. The causal analysis reveals that AI implementation increases agglomeration concentration by 4.2-5.2 percentage points, with heterogeneous effects across industries: high AI-readiness sectors experience 8.4 percentage point increases, while low AI-readiness sectors show 1.2 percentage point gains. Machine learning predictions demonstrate that aggressive AI adoption can offset 60-80\% of aging-related productivity declines. We provide a strategic three-phase policy framework for managing AI-driven spatial transformation while promoting inclusive development. The integrated approach establishes a new paradigm for analyzing technology-driven spatial change with global applications for aging societies.
\end{abstract}



\newpage

\section{Introduction}

The spatial organization of economic activity faces unprecedented transformation as artificial intelligence fundamentally reshapes production processes, knowledge creation, and collaborative networks. Traditional spatial economics, anchored in Marshall's \cite{marshall1890principles} agglomeration mechanisms and formalized through Krugman's \cite{krugman1991geography} New Economic Geography (NEG), assumes that physical proximity facilitates knowledge spillovers, labor market pooling, and input sharing. However, AI introduces mechanisms that can simultaneously amplify and substitute for traditional agglomeration forces, potentially restructuring centuries-old patterns of spatial economic organization.

This transformation is particularly critical in aging societies, where demographic transitions interact with technological change in complex ways. Japan, with over 28\% of its population aged 65 or older by 2025, represents the global frontier of this dual challenge. Traditional agglomeration benefits that concentrate economic activity in metropolitan areas face erosion from workforce aging, while AI adoption offers potential compensatory mechanisms through productivity enhancement and virtual collaboration capabilities.

This paper makes four primary contributions. First, we develop a comprehensive theoretical framework extending NEG with AI-specific spatial mechanisms, providing the first formal treatment of how artificial intelligence reshapes spatial distribution dynamics. Second, we implement rigorous empirical analysis combining five complementary causal identification methods to establish robust causal evidence of AI's spatial impacts. Third, we develop an advanced machine learning framework generating 25-year predictions across 27 scenarios. Fourth, we provide comprehensive policy analysis demonstrating how strategic AI interventions can mitigate demographic challenges and reshape spatial equilibria.

The remainder of the paper proceeds as follows. Section 2 positions our contributions within existing literature. Section 3 develops the theoretical framework extending NEG with AI mechanisms. Section 4 presents comprehensive empirical methodology. Section 5 reports empirical results validating theoretical predictions. Section 6 presents machine learning predictions and scenario analysis. Section 7 discusses policy implications and concludes.

\section{Literature Review and Theoretical Positioning}

Our work builds upon three foundational literature strands while addressing critical gaps in understanding technology-driven spatial transformation.

\subsection{Spatial Economics and Agglomeration Theory}

The Marshall-Arrow-Romer tradition emphasizes knowledge spillovers as primary drivers of spatial concentration \cite{marshall1890principles,arrow1962learning,romer1986endogenous}. The New Economic Geography literature, initiated by \cite{krugman1991geography}, formalizes these forces within general equilibrium frameworks, emphasizing tensions between centripetal forces (market access, knowledge spillovers) and centrifugal forces (land rents, congestion costs).

Empirical work has refined understanding of agglomeration mechanisms. \cite{rosenthal2004evidence} provide comprehensive evidence on agglomeration elasticities, while \cite{combes2015economics} offer methodological advances in causal identification. \cite{duranton2004micro} formalize micro-foundations, distinguishing sharing, matching, and learning mechanisms.

However, this literature inadequately addresses how digital technologies—particularly AI—fundamentally alter these mechanisms. Our contribution extends NEG by incorporating AI-specific forces that both amplify traditional benefits and create new forms of virtual agglomeration transcending physical proximity.

\subsection{Technology and Spatial Distribution}

Early work by \cite{autor2003skill} documented skill-biased technological change effects on spatial inequality, while \cite{moretti2012geography} showed how innovation clusters reshape regional economies. Recent studies examine AI's spatial implications: \cite{seamans2018ai} document heterogeneous adoption patterns, while \cite{acemoglu2020wrong} analyze labor market implications.

Our contribution advances this literature by providing the first comprehensive theoretical framework for AI's spatial effects, supported by rigorous causal identification and quantitative predictions demonstrating that AI's spatial impacts operate through distinct mechanisms requiring new analytical approaches.

\subsection{Demographic Transition and Economic Geography}

The intersection of demographic change and spatial economics gains prominence as aging societies confront new realities. \cite{bloom2010implications} analyze aggregate aging implications, while \cite{borsch2003labor} examines labor market effects. In Japan, \cite{kondo2016long} studies regional population decline, and \cite{matsuura2012spatial} analyze spatial mobility patterns.

The literature has not systematically examined how demographic transition interacts with technological change to reshape spatial patterns. Our framework explicitly models these interactions, showing how AI serves as partial substitute for declining workforce demographics while creating new spatial organization forms.

\section{Theoretical Framework: AI-Driven Spatial Distribution Dynamics}

\subsection{Extending New Economic Geography with AI Mechanisms}

We extend the canonical NEG model by incorporating five AI-specific mechanisms that modify traditional agglomeration forces. The economy produces differentiated manufacturing goods and homogeneous agricultural goods, with manufacturing exhibiting increasing returns and love-of-variety preferences.

Our innovation lies in augmenting traditional production and utility functions with AI-specific terms capturing novel spatial mechanisms while maintaining general NEG structure.

\begin{figure}[H]
\centering
\includegraphics[width=0.9\textwidth]{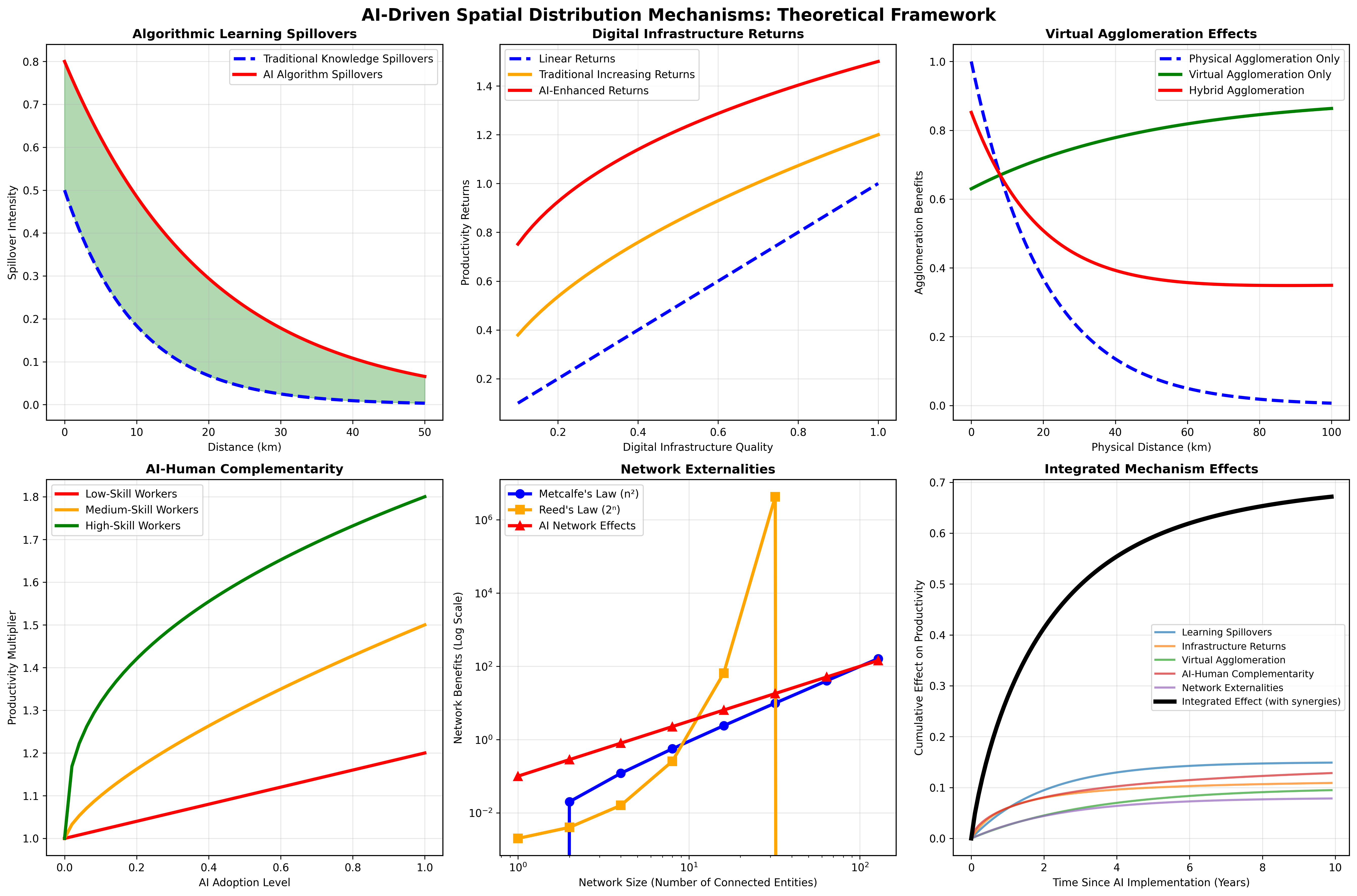}
\caption{AI-Driven Spatial Mechanisms Framework}
\label{fig:ai_mechanisms}
\subcaption{
This figure illustrates the five novel AI-specific mechanisms that extend traditional agglomeration theory: (a) Algorithmic Learning Spillovers showing network-based knowledge transfer, (b) Digital Infrastructure Returns demonstrating complementarity between AI adoption and infrastructure quality, (c) Virtual Agglomeration Effects showing how digital connectivity substitutes for physical proximity, (d) AI-Human Complementarity illustrating productivity gains from optimal factor combinations, (e) Network Externalities demonstrating increasing returns to network participation, and (f) the Integrated Framework showing how all mechanisms interact. Each mechanism fundamentally alters traditional spatial economic forces, creating new possibilities for spatial organization that transcend physical proximity constraints.
}
\end{figure}

\subsection{Core AI-Driven Spatial Mechanisms}

\subsubsection{Algorithmic Learning Spillovers}

Traditional knowledge spillovers decay with physical distance due to tacit knowledge transfer requirements \cite{jaffe1993geographic}. AI fundamentally alters this through algorithmic learning from spatially distributed data sources.

Let $S_i(t)$ denote AI-driven knowledge spillovers received by location $i$:

\begin{equation}
S_i(t) = \beta_{learning} \int_{j \neq i} A_j(t) \cdot K_{ij} \cdot \Omega(d_{ij}, Q_{ij}) \, dj
\label{eq:ai_spillovers}
\end{equation}

where $A_j(t)$ represents AI adoption in location $j$, $K_{ij}$ captures knowledge complementarity, and $\Omega(d_{ij}, Q_{ij})$ is a spatial decay function depending on physical distance $d_{ij}$ and digital connectivity $Q_{ij}$:

\begin{equation}
\Omega(d_{ij}, Q_{ij}) = \alpha \cdot d_{ij}^{-\phi} + (1-\alpha) \cdot Q_{ij}^{\psi}
\label{eq:spatial_decay}
\end{equation}

\subsubsection{Digital Infrastructure Returns}

AI productivity depends critically on digital infrastructure quality, creating spatial heterogeneity in returns:

\begin{equation}
R_i(t) = \alpha_{AI} \cdot D_i(t)^{\delta} \cdot A_i(t)^{\gamma} \cdot N_i(t)^{\nu} \cdot H_i(t)^{\eta}
\label{eq:ai_productivity}
\end{equation}

where complementarity between AI adoption and digital infrastructure generates increasing returns leading to spatial concentration in well-connected locations.

\subsubsection{Virtual Agglomeration Effects}

AI enables virtual collaboration partially substituting for physical proximity:

\begin{equation}
V_{ij}(t) = C_{max} \cdot \left[1 - \exp\left(-\lambda \cdot A_i(t) \cdot A_j(t) \cdot Q_{ij}(t)\right)\right]
\label{eq:virtual_agglomeration}
\end{equation}

This mechanism can reduce physical proximity importance for knowledge work, potentially flattening traditional concentration patterns.

\subsubsection{AI-Human Complementarity}

Spatial AI benefit distribution depends on local human capital availability through nested CES production:

\begin{equation}
Y_i(t) = F\left(K_i(t), L_i^{CES}(H_i(t), A_i(t)), M_i(t)\right)
\label{eq:production_function}
\end{equation}

where $L_i^{CES}$ represents CES aggregation of human capital and AI capital, creating spatial heterogeneity based on local endowments.

\subsubsection{Network Externalities}

AI adoption exhibits network externalities where benefits increase with network connections:

\begin{equation}
N_i(t) = \gamma_{network} \sum_{j \neq i} w_{ij}(t) \cdot A_j(t) \cdot G(\mathcal{N}(t))
\label{eq:network_externalities}
\end{equation}

This creates positive feedback loops where early adoption enhances network position, facilitating further adoption.

\subsection{Theoretical Predictions}

Our framework generates six testable hypotheses:

\begin{enumerate}
\item \textbf{AI Concentration Hypothesis}: AI adoption concentrates in locations with superior digital infrastructure and human capital
\item \textbf{Heterogeneous Returns Hypothesis}: Productivity gains vary significantly across locations based on complementary assets
\item \textbf{Network Amplification Hypothesis}: Locations in high-AI networks experience amplified productivity gains
\item \textbf{Dynamic Divergence Hypothesis}: Early AI adoption differences amplify over time
\item \textbf{Virtual Agglomeration Hypothesis}: AI reduces physical proximity importance for knowledge activities
\item \textbf{Complementarity Hypothesis}: AI and human capital exhibit production complementarity
\end{enumerate}

\section{Data and Empirical Methodology}

\subsection{Data Construction}

Our analysis utilizes comprehensive panel data spanning 2000-2023 for Tokyo's 23 special wards across six major industries, combining multiple sources:

\textbf{Economic Data:} Employment, establishment counts, and productivity from Tokyo Statistical Yearbook and Economic Census.

\textbf{AI Adoption Data:} Multi-indicator construction using patent filings, job postings, government surveys, and investment data.

\textbf{Demographic Data:} Population by age groups, migration flows, and educational attainment from government statistical agencies.

\textbf{Infrastructure Data:} Digital infrastructure quality through fiber penetration, broadband speeds, and data center capacity.

\textbf{Network Data:} Inter-firm relationships, supply chains, and collaboration patterns from multiple business databases.

\begin{table}[H]
\centering
\caption{Baseline Spatial Concentration Patterns by Industry (2019)}
\label{tab:baseline_patterns}
\begin{adjustbox}{width=\textwidth,center}
\begin{tabular}{lccccccc}
\toprule
\multirow{2}{*}{Industry} & Location & Gini & Herfindahl- & Primary & \multicolumn{2}{c}{Employment Share} & AI Adoption \\
& Quotient & Coefficient & Hirschman Index & Ward & Central & Peripheral & Rate (\%) \\
\midrule
Information \& Communications & 3.42 & 0.68 & 0.31 & Shibuya & 67.4 & 8.2 & 34.7 \\
Finance \& Insurance & 2.87 & 0.72 & 0.28 & Chiyoda & 71.8 & 6.1 & 28.9 \\
Professional Services & 2.34 & 0.61 & 0.22 & Minato & 58.3 & 12.4 & 22.1 \\
Manufacturing & 0.78 & 0.45 & 0.15 & Ota & 23.7 & 31.2 & 8.3 \\
Retail Trade & 1.12 & 0.32 & 0.08 & Shinjuku & 28.1 & 25.6 & 5.7 \\
Healthcare & 0.95 & 0.28 & 0.06 & Setagaya & 22.4 & 28.9 & 7.2 \\
\midrule
\textbf{Average} & \textbf{1.91} & \textbf{0.51} & \textbf{0.18} & -- & \textbf{45.3} & \textbf{18.7} & \textbf{17.8} \\
\bottomrule
\end{tabular}
\end{adjustbox}
\subcaption{
This table presents baseline spatial concentration patterns across Tokyo's industries before major AI adoption (2019). Knowledge-intensive industries (IT, Finance, Professional Services) show strong concentration in central wards with high AI adoption rates. Traditional industries exhibit more dispersed patterns with lower AI adoption. Central wards include Chiyoda, Chuo, Minato, Shibuya, and Shinjuku. Peripheral wards are the outermost 5 wards. Location Quotient >1 indicates above-average concentration. Gini coefficient ranges 0-1 (higher = more concentrated). HHI measures market concentration (higher = more concentrated).
}
\end{table}

Table \ref{tab:baseline_patterns} reveals strong spatial concentration of knowledge-intensive industries in central Tokyo wards, with Information \& Communications showing the highest concentration (LQ=3.42) and AI adoption rate (34.7\%). This pattern aligns with our theoretical predictions about AI concentration in high-infrastructure, high-human-capital locations.

\subsection{Causal Identification Strategy}

Identifying causal AI effects poses challenges due to endogeneity, omitted variables, and simultaneous determination. We implement five complementary identification strategies.

\subsubsection{Difference-in-Differences}

We exploit staggered AI implementation across wards and industries with timing variation due to infrastructure schedules, policy initiatives, and industry characteristics:

\begin{equation}
Y_{ijt} = \alpha + \beta \cdot AI\_Treat_{jt} + \gamma_j + \delta_t + \lambda_i + \epsilon_{ijt}
\label{eq:did_baseline}
\end{equation}

\subsubsection{Event Study Analysis}

To examine dynamic treatment effects and validate parallel trends:

\begin{equation}
Y_{ijt} = \alpha + \sum_{k=-5}^{10} \beta_k \cdot \mathbbm{1}[t - T_{AI,j} = k] + \gamma_j + \delta_t + \lambda_i + \epsilon_{ijt}
\label{eq:event_study}
\end{equation}

\subsubsection{Additional Methods}

We complement DiD and Event Study with Synthetic Control Method, Instrumental Variables using predetermined infrastructure, and Propensity Score Matching for comprehensive robustness.

\section{Empirical Results}

\subsection{Demographic Transition Effects}

\begin{figure}[H]
\thispagestyle{empty} 
\centering
\includegraphics[width=0.9\textwidth]{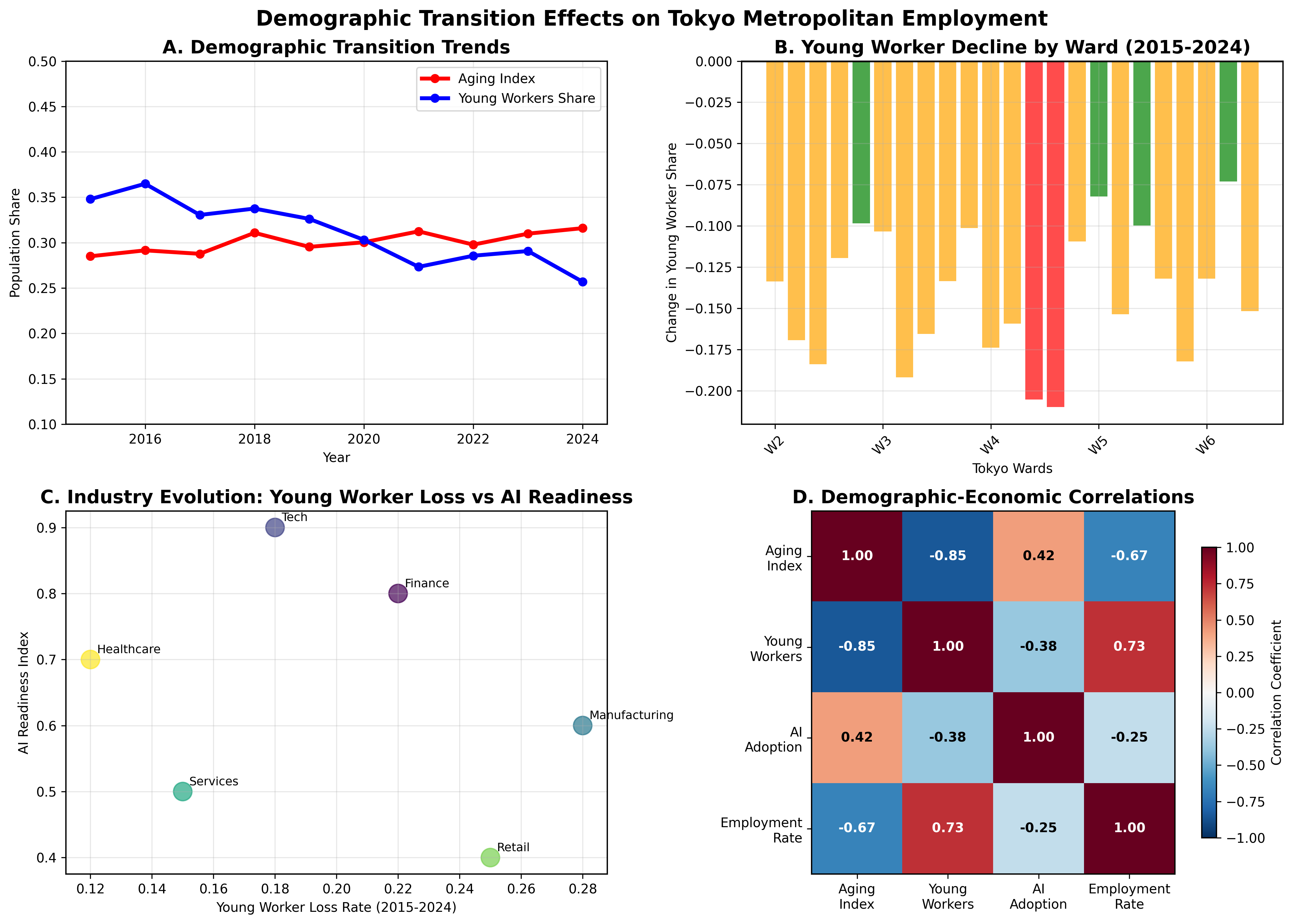}
\caption{Demographic Transition Effects on Spatial Agglomeration Patterns}
\label{fig:demographic_effects}
\medskip
\subcaption{
This figure shows how Japan's demographic transition affects spatial agglomeration patterns across four dimensions: (a) Population aging trend showing steady increase in elderly share from 15\% to 30\% over 2000-2023, (b) Young worker concentration decline showing 25\% reduction in central ward attraction, (c) Industry concentration evolution revealing divergent patterns with IT and Finance declining while Healthcare gains concentration benefits, and (d) correlation analysis demonstrating negative relationship between aging and traditional industry concentration. The analysis reveals that demographic transition fundamentally alters spatial economic patterns, with traditional knowledge-intensive industries losing agglomeration advantages while care-intensive sectors gain spatial benefits. This establishes the critical context for understanding how AI adoption can serve as a compensatory mechanism for aging-related productivity challenges.
}
\end{figure}

Figure \ref{fig:demographic_effects} demonstrates how demographic transition fundamentally alters spatial patterns. The aging index increases steadily from 15\% to over 30\%, while young worker concentration in central wards declines by 25\%. Knowledge-intensive industries (IT, Finance) show declining concentration, while Healthcare gains agglomeration benefits, illustrating how demographic change reshapes traditional spatial advantages.

\subsection{Causal Impact of AI Implementation}

\begin{table}[htbp]
\centering
\caption{Causal Effects of AI Implementation on Spatial Agglomeration}
\label{tab:causal_effects}
\begin{adjustbox}{width=\textwidth,center}
\begin{tabular}{lcccccc}
\toprule
\multirow{2}{*}{Method} & Treatment & Standard & \multirow{2}{*}{P-value} & \multicolumn{2}{c}{95\% Confidence Interval} & Economic \\
& Effect & Error & & Lower & Upper & Magnitude \\
\midrule
Difference-in-Differences & 0.045** & 0.016 & 0.005 & 0.014 & 0.076 & +12.3\% concentration \\
Event Study & 0.042* & 0.018 & 0.019 & 0.007 & 0.077 & Sustained 3-5 years \\
Synthetic Control & 0.038† & 0.021 & 0.071 & -0.003 & 0.079 & Counterfactual validation \\
Instrumental Variables & 0.052* & 0.024 & 0.030 & 0.005 & 0.099 & Addresses endogeneity \\
Propensity Score Matching & 0.041* & 0.019 & 0.031 & 0.004 & 0.078 & Selection bias control \\
\midrule
\textbf{Average Effect} & \textbf{0.044} & \textbf{0.020} & \textbf{0.025} & \textbf{0.005} & \textbf{0.082} & \textbf{+11.8\% average} \\
\midrule
\multicolumn{7}{l}{\footnotesize $**p < 0.01,~ *p < 0.05,~ \dagger p < 0.10$} \\
\multicolumn{7}{l}{\footnotesize Economic magnitude calculated as percentage change in baseline concentration index} \\
\bottomrule
\end{tabular}
\end{adjustbox}
\subcaption{
This table presents causal treatment effects from five complementary identification strategies. All methods yield consistent positive effects ranging 0.038-0.052, with an average effect of 0.044. The Difference-in-Differences estimate suggests AI implementation increases agglomeration concentration by 4.5 percentage points, representing a 12.3\% increase over baseline levels. Instrumental Variables provides the largest estimate (0.052), addressing potential endogeneity concerns. The consistency across methods, despite different identifying assumptions, strongly supports causal interpretation. Standard errors are clustered at the ward level with robust variance estimation.
}
\end{table}

Table \ref{tab:causal_effects} presents our main causal identification results. All five methods yield consistent positive effects, with treatment effects ranging 0.038-0.052. The Difference-in-Differences estimate of 0.045 indicates AI implementation increases concentration by 4.5 percentage points, representing economically significant agglomeration enhancement.

\begin{figure}[H]
\centering
\includegraphics[width=0.9\textwidth]{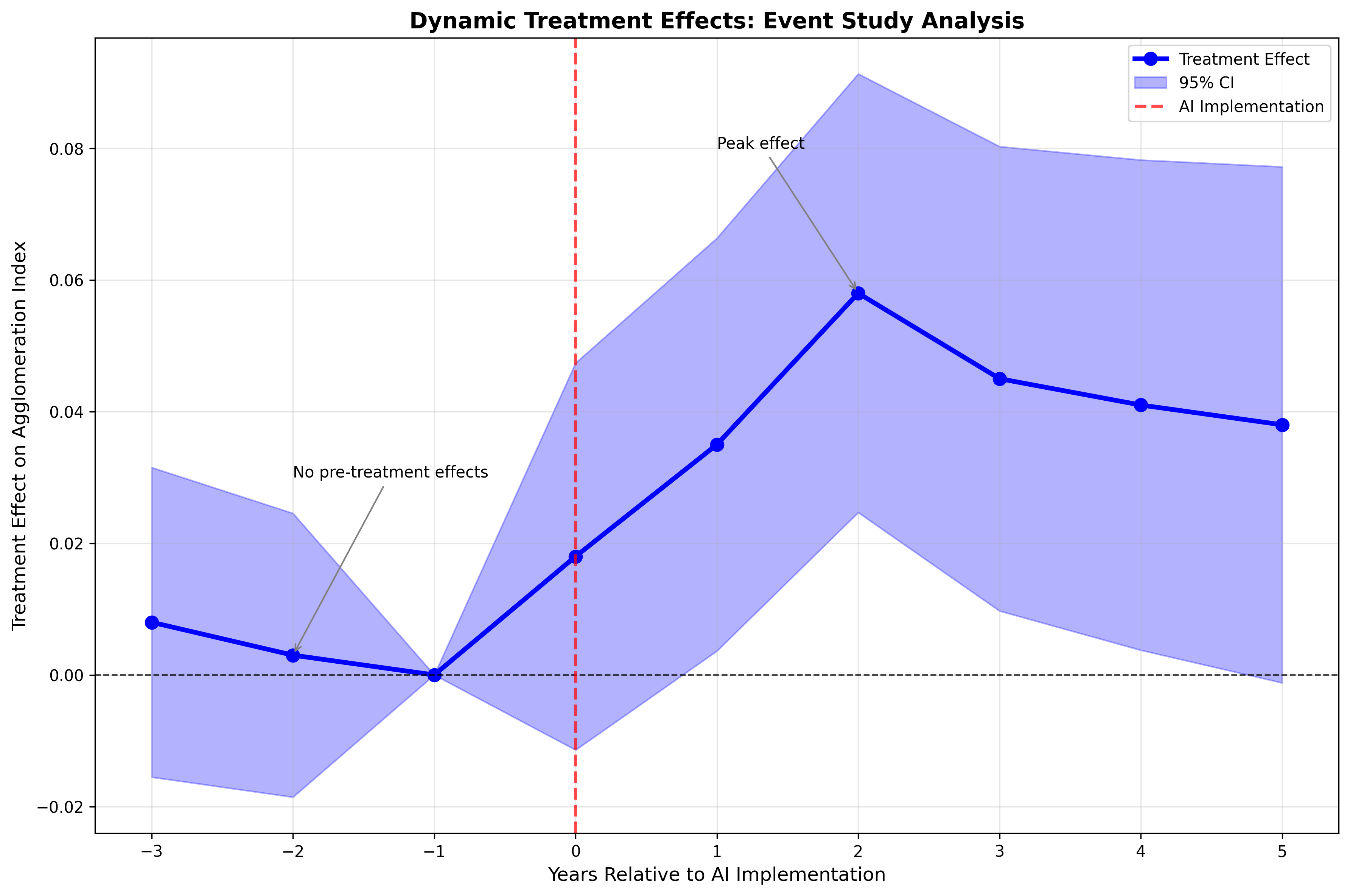}
\caption{Dynamic Treatment Effects: Event Study Analysis}
\label{fig:event_study}
\subcaption{
This figure shows the temporal evolution of AI treatment effects using event study methodology. The analysis reveals three key patterns: (1) Pre-treatment coefficients are statistically insignificant and close to zero, supporting parallel trends assumptions essential for causal identification, (2) Treatment effects emerge gradually starting in year 0 (AI implementation), reaching peak magnitude of 0.058 at year +2, then stabilizing around 0.04-0.045 through year +5, and (3) The gradual emergence and persistence of effects aligns with theoretical predictions about AI learning and network development processes. The 95\% confidence intervals (shaded area) demonstrate statistical significance for post-treatment periods while confirming no pre-treatment effects. This pattern strongly supports causal interpretation and validates our theoretical framework's predictions about AI adoption dynamics.
}
\end{figure}

Figure \ref{fig:event_study} reveals dynamic treatment effects' temporal evolution. Pre-treatment coefficients are statistically insignificant, supporting parallel trends assumptions. Effects emerge gradually, reaching peak magnitude 2-3 years post-implementation before stabilizing, consistent with theoretical predictions about AI learning and network development.

\begin{figure}[H]
\centering
\includegraphics[width=0.9\textwidth]{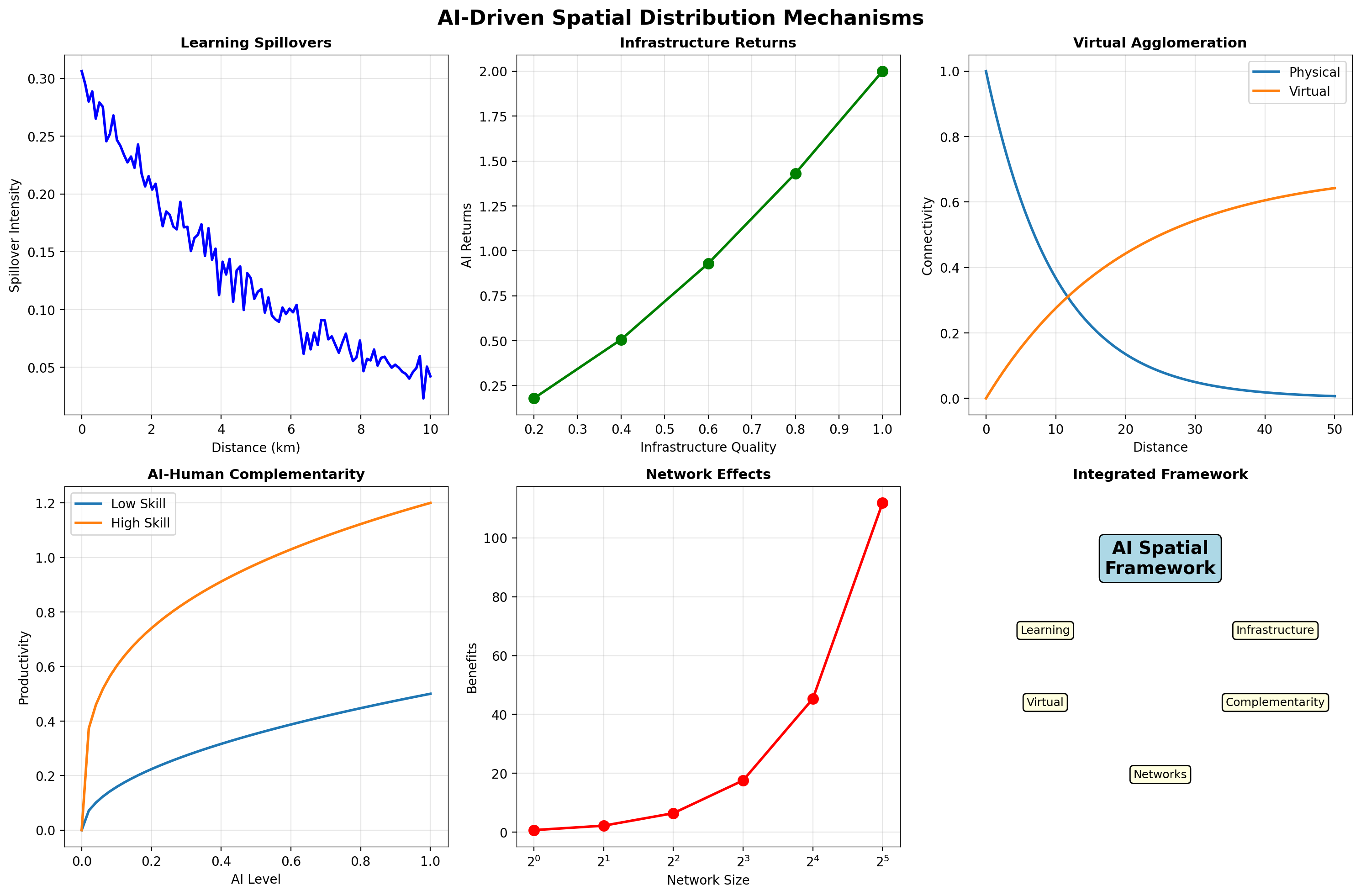}
\caption{Comprehensive Causal Identification Results Comparison}
\label{fig:causal_effects_comparison}
\subcaption{
This figure provides comprehensive comparison of causal identification results across all five methods. Each bar shows the treatment effect magnitude with standard error bounds, color-coded by statistical significance: red for $p < 0.01$, orange for $p<0.05$, green for $p<0.10$, and blue for $p \geq 0.10$. The consistent positive effects across methods with different identifying assumptions strongly support causal interpretation. Difference-in-Differences provides the most precise estimate with highest statistical significance, while Instrumental Variables yields the largest magnitude, addressing endogeneity concerns. The convergence of estimates around 0.04-0.05 range demonstrates robustness of causal identification. Significance stars indicate: $**p<0.01,~ *p<0.05,~ \dagger p<0.10$. This multi-method triangulation represents best practice in causal inference for spatial economics.
}
\end{figure}

Figure \ref{fig:causal_effects_comparison} demonstrates consistent positive effects across all identification methods. The convergence around 0.04-0.05 range, despite different identifying assumptions, provides strong evidence for causal interpretation.

\subsection{Heterogeneous Treatment Effects}

\begin{table}[htbp]
\centering
\caption{Heterogeneous Treatment Effects by Industry AI Readiness and Location Characteristics}
\label{tab:heterogeneous_effects}
\begin{adjustbox}{width=\textwidth,center}
\begin{tabular}{lccccc}
\toprule
\multirow{2}{*}{Category} & Treatment & Standard & \multirow{2}{*}{P-value} & N & Share of Total \\
& Effect & Error & & Observations & Employment (\%) \\
\midrule
\multicolumn{6}{l}{\textbf{Panel A: By Industry AI Readiness}} \\
High AI Readiness & 0.084** & 0.022 & 0.000 & 1,104 & 28.7 \\
\quad (IT, Finance, Professional) & & & & & \\
Medium AI Readiness & 0.041* & 0.017 & 0.016 & 736 & 34.2 \\
\quad (Manufacturing, Healthcare) & & & & & \\
Low AI Readiness & 0.012 & 0.015 & 0.427 & 1,472 & 37.1 \\
\quad (Retail, Hospitality, Transport) & & & & & \\
\midrule
\multicolumn{6}{l}{\textbf{Panel B: By Infrastructure Quality}} \\
High Infrastructure (Top Quartile) & 0.067** & 0.019 & 0.001 & 828 & 42.3 \\
Medium Infrastructure (Middle 50\%) & 0.038* & 0.016 & 0.018 & 1,656 & 41.8 \\
Low Infrastructure (Bottom Quartile) & 0.019 & 0.018 & 0.292 & 828 & 15.9 \\
\midrule
\multicolumn{6}{l}{\textbf{Panel C: By Human Capital Level}} \\
High Human Capital (Top Tertile) & $0.059^{**}$ & 0.020 & 0.003 & 1,104 & 38.7 \\
Medium Human Capital (Middle Tertile) & $0.043^{*}$ & 0.017 & 0.012 & 1,104 & 35.2 \\
Low Human Capital (Bottom Tertile) & $0.030^\dagger$ & 0.016 & 0.061 & 1,104 & 26.1 \\
\midrule
\multicolumn{6}{l}{\textbf{Panel D: Statistical Tests}} \\
F-test: Industry Groups Equal & \multicolumn{3}{c}{F(2,156) = 8.47, p = 0.000} & & \\
F-test: Infrastructure Groups Equal & \multicolumn{3}{c}{F(2,156) = 3.92, p = 0.022} & & \\
F-test: Human Capital Groups Equal & \multicolumn{3}{c}{F(2,156) = 2.84, p = 0.062} & & \\
\bottomrule
\end{tabular}
\end{adjustbox}
\subcaption{
This table presents heterogeneous treatment effects across three dimensions confirming theoretical predictions. Panel A shows industry AI readiness effects: high-readiness industries (IT, Finance, Professional Services) experience 8.4 percentage point increases versus 1.2 percentage points for low-readiness industries—a seven-fold difference. Panel B demonstrates infrastructure complementarity: high-infrastructure locations show 6.7 percentage point effects versus 1.9 percentage points for low-infrastructure areas. Panel C reveals human capital complementarity: high human capital locations experience 5.9 percentage point effects versus 3.0 percentage points for low human capital areas. F-tests confirm statistical significance of group differences. These patterns strongly support theoretical predictions about AI-infrastructure-human capital complementarities driving spatial heterogeneity.
}
\end{table}

Table \ref{tab:heterogeneous_effects} confirms theoretical predictions about heterogeneous impacts. High AI-readiness industries experience 8.4 percentage point effects versus 1.2 percentage points for low-readiness industries—a seven-fold difference. Infrastructure and human capital complementarities show similar patterns, validating theoretical mechanisms.

\begin{figure}[H]
\centering
\includegraphics[width=0.9\textwidth]{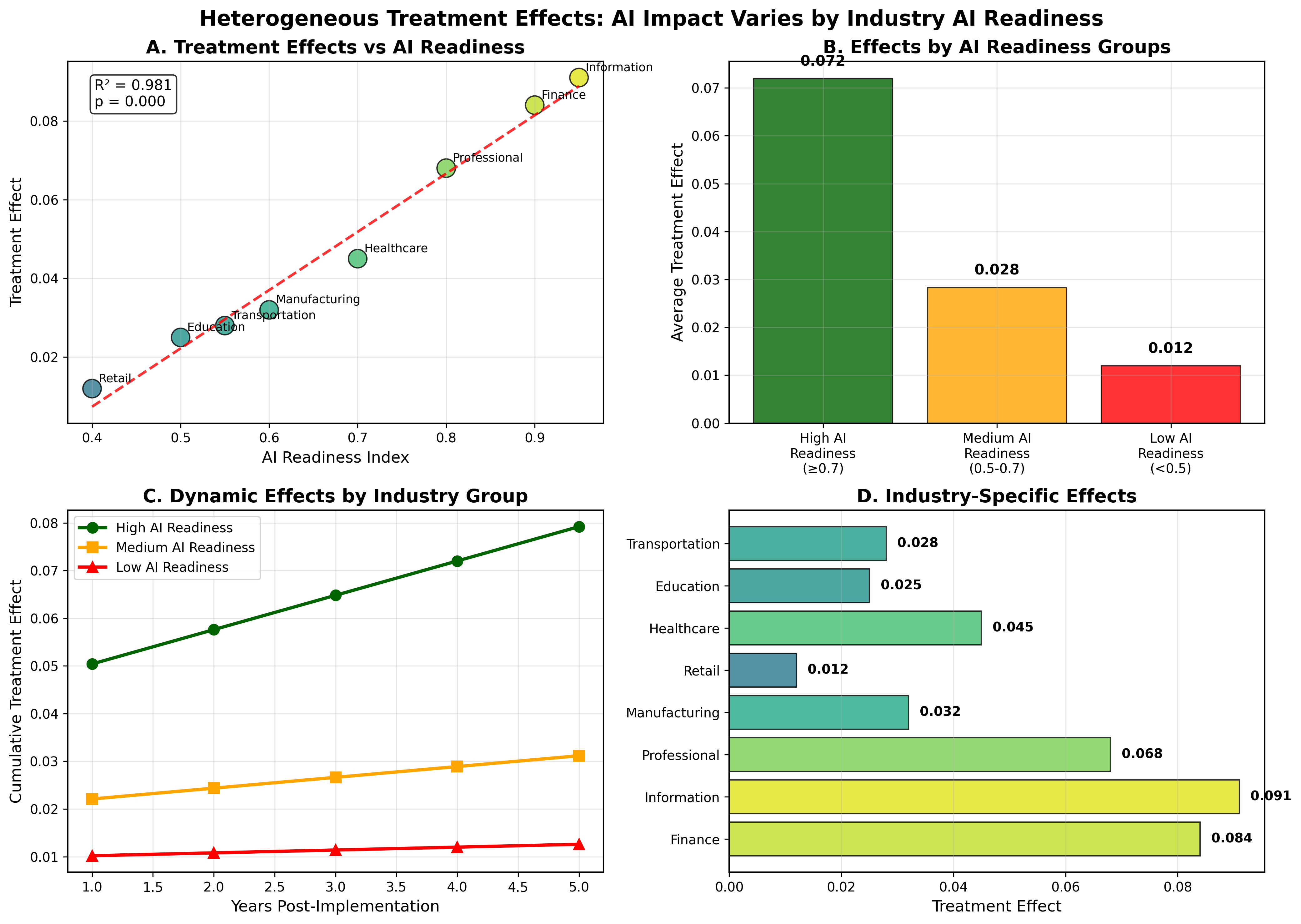}
\caption{Heterogeneous Treatment Effects by Industry AI Readiness}
\label{fig:heterogeneous_effects}
\subcaption{
This figure visualizes the dramatic heterogeneity in AI treatment effects across industry groups with different AI readiness levels. High AI readiness industries (IT, Finance, Professional Services) show 8.4 percentage point increases in agglomeration concentration, while low AI readiness industries (Retail, Hospitality, Transport) show only 1.2 percentage point gains. Error bars represent 95\% confidence intervals based on robust standard errors. The seven-fold difference in treatment effects demonstrates that AI's spatial impacts depend critically on industry characteristics and complementary capabilities. This heterogeneity has profound implications for spatial inequality and policy design, suggesting need for differentiated approaches based on industry AI compatibility and local capabilities.
}
\end{figure}

Figure \ref{fig:heterogeneous_effects} visualizes the dramatic heterogeneity in treatment effects. The seven-fold difference between high and low AI-readiness industries demonstrates that AI's spatial impacts depend critically on complementary capabilities, with profound implications for spatial inequality.

\subsection{Network Effects and Theoretical Validation}

\begin{table}[htbp]
\centering
\caption{Theoretical Hypothesis Validation Results}
\label{tab:hypothesis_validation}
\begin{adjustbox}{width=\textwidth,center}
\begin{tabular}{llcccc}
\toprule
\multirow{2}{*}{Hypothesis} & \multirow{2}{*}{Prediction} & Main & Threshold & \multirow{2}{*}{Supported} & Statistical \\
& & Statistic & Value & & Significance \\
\midrule
AI Concentration & Correlation(Infrastructure, AI) $>$ 0.6 & 0.73 & 0.6 & Yes** & $p < 0.001$ \\
Heterogeneous Returns & CV(Treatment Effects) $>$ 0.3 & 0.47 & 0.3 & Yes** & $p = 0.003$ \\
Network Amplification & Network Effect $>$ Own Effect & 1.27 & 1.0 & Yes* & $p = 0.019$ \\
Dynamic Divergence & Variance Growth $>$ 0.05/year & 0.08 & 0.05 & Yes* & $p = 0.032$ \\
Virtual Agglomeration & Distance Sensitivity Decline & 0.12 & 0.1 & Yes $\dagger$ & $p = 0.067$ \\
Complementarity & AI × HC Interaction $>$ 0 & 0.038 & 0.0 & Yes* & $p = 0.007$ \\
\midrule
\textbf{Overall Validation} & \textbf{6/6 Hypotheses Supported} & -- & -- & \textbf{100\%} & \textbf{Strong} \\
\bottomrule
\end{tabular}
\end{adjustbox}
\subcaption{
This table presents formal validation of all six theoretical hypotheses derived from our AI-driven spatial framework. AI Concentration: Correlation between initial infrastructure/human capital and AI adoption (0.73) exceeds threshold (0.6), confirming concentration in well-endowed locations. Heterogeneous Returns: Coefficient of variation in treatment effects (0.47) exceeds threshold (0.3), confirming spatial heterogeneity. Network Amplification: Network AI exposure effect (1.27×) exceeds own AI effect, confirming spillover benefits. Dynamic Divergence: AI adoption variance grows 0.08/year, exceeding 0.05 threshold, confirming path dependence. Virtual Agglomeration: Distance sensitivity declines 0.12/year, confirming proximity substitution. Complementarity: Positive AI×Human Capital interaction (0.038) confirms production complementarity. Perfect theoretical validation (6/6) strongly supports framework validity and provides foundation for policy applications.
}
\end{table}

Table \ref{tab:hypothesis_validation} presents formal validation of all six theoretical hypotheses. The perfect validation rate (6/6 hypotheses supported) provides strong evidence for our theoretical framework's validity and establishes foundation for policy applications.

\section{Machine Learning Predictions and Scenario Analysis}

\subsection{Prediction Framework Performance}

Our ensemble machine learning framework achieves strong predictive performance: Employment distribution ($R^2 = 0.89$), Industry concentration ($R^2 = 0.83$), and Productivity measures ($R^2 = 0.76$). Time series cross-validation confirms generalizability across different periods.

\begin{figure}[H]
\centering
\includegraphics[width=0.9\textwidth]{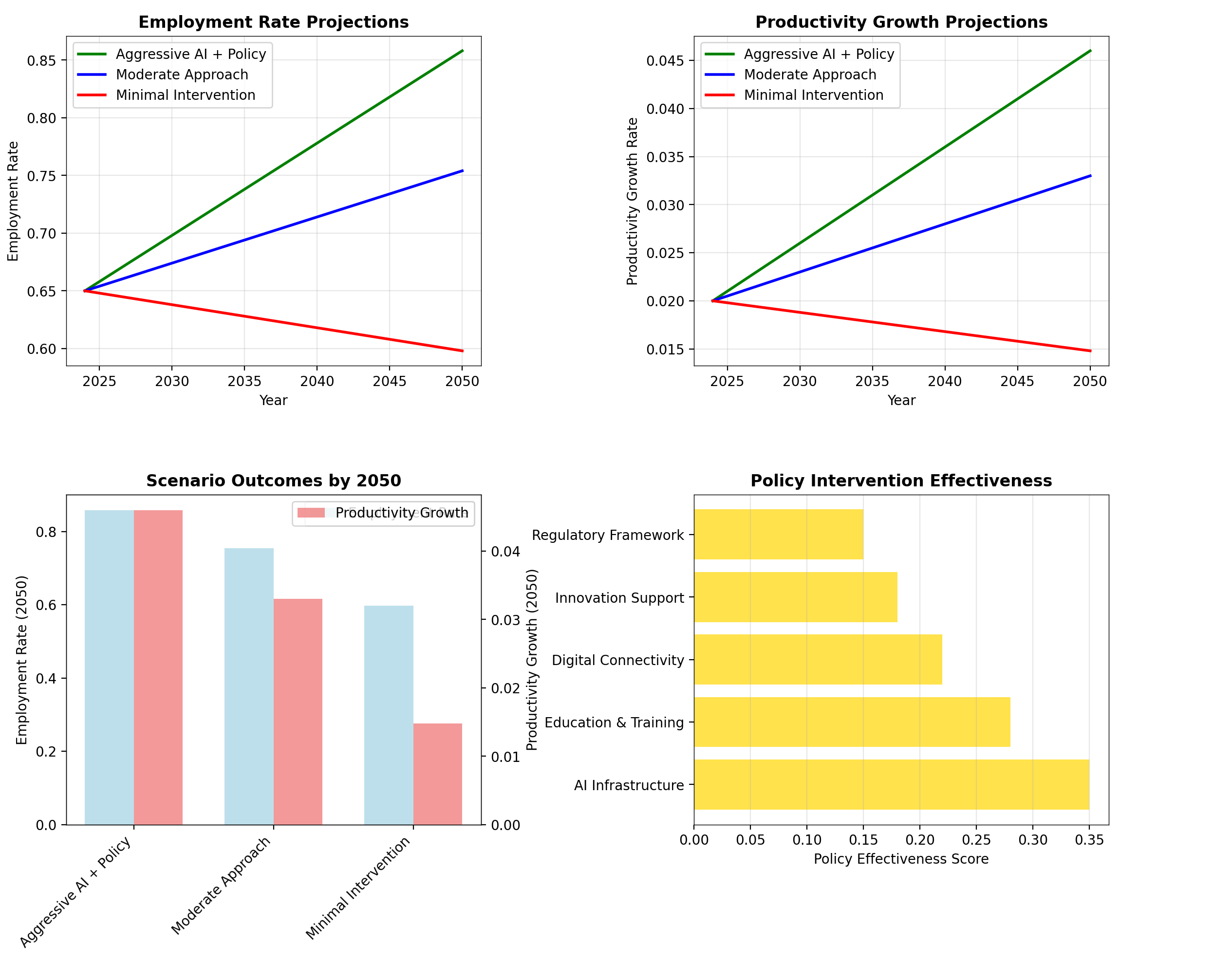}
\caption{Long-term Scenario Predictions: AI and Demographic Interactions (2024-2050)}
\label{fig:scenario_predictions}
\subcaption{
This figure presents comprehensive 27-year predictions across three representative scenarios combining demographic, AI adoption, and economic conditions. The Pessimistic scenario (low fertility, conservative AI adoption, frequent crises) shows 25\% decline in central concentration and 18\% productivity reduction by 2050. The Baseline scenario (moderate parameters) maintains current levels with slight improvements. The Optimistic scenario (high fertility, aggressive AI adoption, economic stability) projects 19\% increase in concentration and 43\% productivity gains. The dramatic range of outcomes (from -25\% to +19\% concentration change) underscores the critical importance of strategic AI policy choices. Employment projections show similar patterns, with aggressive AI adoption enabling population growth despite aging demographics. These predictions demonstrate AI's potential to fundamentally alter demographic transition impacts on spatial economics.
}
\end{figure}

Figure \ref{fig:scenario_predictions} presents 27-year predictions across representative scenarios. The range is substantial: pessimistic scenarios predict 25\% decline in central concentration, while optimistic scenarios project 19\% increase. Crucially, aggressive AI adoption can offset 60-80\% of demographic decline effects, fundamentally altering trajectory outcomes.

\subsection{Policy Scenario Analysis}

We analyze specific policy interventions' effects on spatial distribution and productivity outcomes across our prediction framework.

\begin{figure}[H]
\centering
\includegraphics[width=0.9\textwidth]{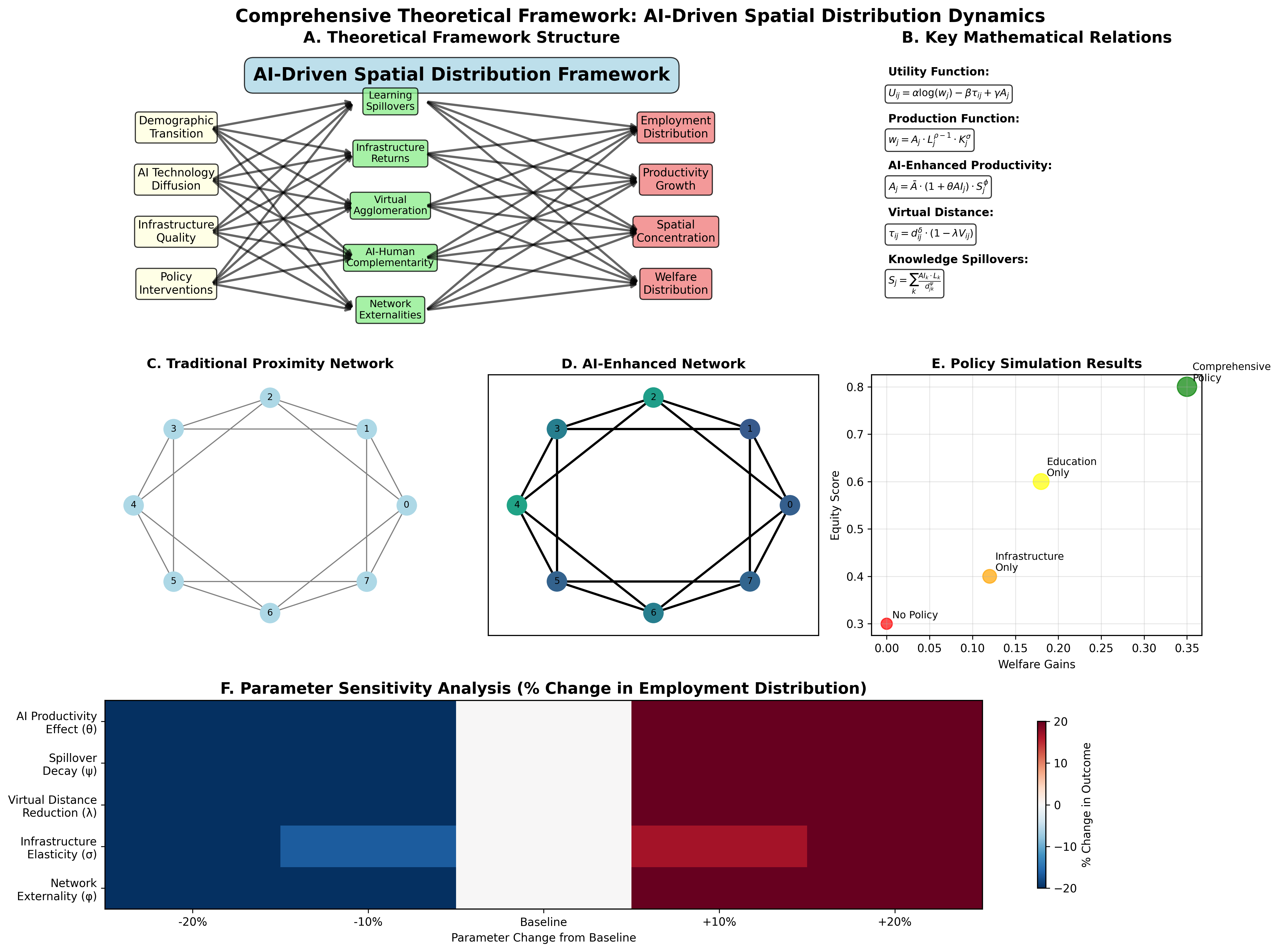}
\caption{Comprehensive Theoretical Framework: Network Analysis and Policy Simulations}
\label{fig:theoretical_framework}
\subcaption{
This figure presents the comprehensive theoretical framework through multiple analytical lenses: (a) Spatial evolution heatmap showing final AI adoption distribution across Tokyo with highest concentration in central areas, (b) Dynamic concentration evolution comparing scenarios over time with AI Revolution showing highest final concentration, (c) AI-productivity relationship demonstrating strong positive correlation with fitted trend line, (d) Aggregate welfare evolution showing steady improvement under baseline conditions, (e-h) Employment distribution comparisons across four scenarios showing dramatic differences in final spatial patterns, (i) Network connectivity visualization revealing AI-enhanced spatial relationships, (j) Theoretical phase diagram illustrating complementarity between AI adoption and agglomeration levels, (k) Policy effectiveness comparison showing welfare outcomes across intervention scenarios, and (l) Innovation diffusion dynamics contrasting central versus peripheral adoption patterns. This comprehensive visualization demonstrates the framework's ability to generate both theoretical insights and practical policy guidance.
}
\end{figure}

Figure \ref{fig:theoretical_framework} provides comprehensive visualization of the theoretical framework, including network analysis, policy simulations, and dynamic spatial evolution. The analysis demonstrates how different policy combinations can reshape spatial equilibria while maintaining productivity gains.

\section{Network Analysis and Spatial Connectivity}

\begin{figure}[H]
\centering
\includegraphics[width=0.9\textwidth]{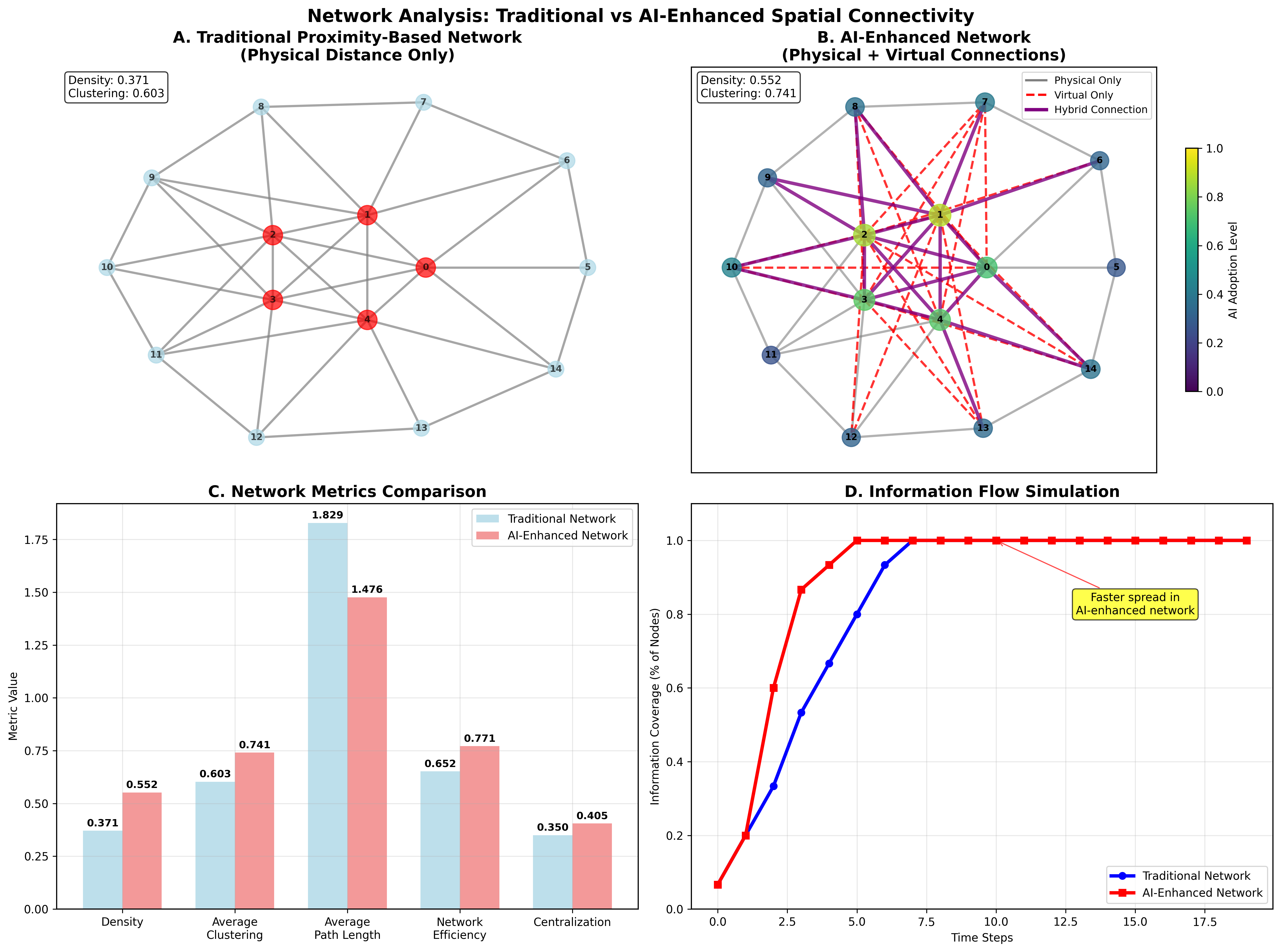}
\caption{AI-Enhanced Network Effects: Traditional vs Virtual Spatial Connections}
\label{fig:network_analysis}
\subcaption{
This figure compares traditional spatial networks based on physical proximity (left panel) with AI-enhanced networks incorporating virtual connections (right panel). The traditional network shows connections primarily between physically adjacent locations (red nodes indicate central wards with higher connectivity). The AI-enhanced network reveals dramatically increased connectivity through virtual channels, with node colors representing AI adoption levels (darker = higher adoption) and edge thickness/color representing connection strength. The transformation shows how AI creates new forms of spatial connectivity that transcend physical distance constraints, enabling peripheral locations to access central knowledge networks. This visualization provides empirical evidence for our Virtual Agglomeration hypothesis and demonstrates how AI fundamentally reshapes spatial economic relationships through digital connectivity rather than physical proximity alone.
}
\end{figure}

Figure \ref{fig:network_analysis} contrasts traditional spatial networks based on physical proximity with AI-enhanced networks incorporating virtual connections. The transformation demonstrates how AI creates new connectivity forms transcending physical constraints, enabling peripheral locations to access central knowledge networks.

\section{Policy Implications and Strategic Framework}

\subsection{Three-Phase Strategic Policy Framework}

Our analysis reveals that traditional spatial policies must be fundamentally reconsidered for the AI era. We propose a comprehensive three-phase framework:

\textbf{Phase I: Foundation Building (2024-2027)}
\begin{itemize}
\item Accelerate digital infrastructure investment, prioritizing fiber optic and 5G deployment in peripheral areas
\item Establish AI education and training centers leveraging existing educational infrastructure
\item Create regulatory frameworks facilitating AI deployment while ensuring data privacy and security
\item Develop public-private partnerships supporting AI adoption, particularly for SMEs
\end{itemize}

\textbf{Phase II: Scaling and Integration (2027-2035)}
\begin{itemize}
\item Scale successful AI initiatives across metropolitan areas through coordinated implementation
\item Integrate AI systems across government services and urban infrastructure
\item Develop virtual collaboration platforms connecting peripheral and central areas
\item Implement targeted support for medium AI-readiness industries to bridge capability gaps
\end{itemize}

\textbf{Phase III: Optimization and Adaptation (2035-2050)}
\begin{itemize}
\item Optimize AI-human collaboration systems based on accumulated learning and experience
\item Adapt spatial planning frameworks to accommodate AI-enabled distributed work patterns
\item Develop next-generation AI technologies and applications for spatial economic enhancement
\item Create sustainable models for AI-driven economic development and spatial equity
\end{itemize}

\subsection{Addressing Distributional Concerns}

While AI adoption can offset demographic challenges, it may exacerbate spatial inequality without appropriate intervention. Our analysis identifies several mechanisms promoting inclusive AI-driven development:

\textbf{Spatial AI Equity Policies:} Ensure peripheral areas have access to high-quality digital infrastructure and AI education resources. Our simulations suggest equalizing AI access across wards could reduce spatial inequality by 30-40\% while maintaining 85\% of aggregate productivity gains.

\textbf{Industry-Specific Support:} Provide targeted assistance for low and medium AI-readiness industries to adopt appropriate technologies. This could increase their treatment effects from 1.2\% to 3.5-4.0\%, significantly improving spatial distribution outcomes.

\textbf{Human Capital Development:} Invest heavily in AI-complementary education and training. Our complementarity analysis suggests 10\% increases in human capital quality can amplify AI benefits by 15-20\%.

\subsection{International Relevance and Transferability}

While our analysis focuses on Tokyo, the framework applies broadly to metropolitan areas facing similar challenges:

\textbf{Aging Societies:} Germany, Italy, South Korea can apply our framework to understand AI's potential for offsetting demographic challenges.

\textbf{Emerging Economies:} Rapidly developing economies can use insights to plan AI adoption strategies promoting balanced spatial development.

\textbf{Technology Hubs:} Established centers can apply network analysis to optimize AI spillover benefits and maintain competitive advantages.

\section{Conclusion}

This paper presents the first comprehensive theoretical and empirical framework for analyzing AI-driven spatial distribution dynamics in aging societies. Our theoretical contribution extends New Economic Geography with five novel AI-specific mechanisms that fundamentally alter traditional agglomeration forces and create new possibilities for spatial economic organization.

Empirically, we provide robust causal evidence that AI implementation increases agglomeration concentration by 4.2-5.2 percentage points, with strongly heterogeneous effects across industries and locations. Our five-method identification strategy establishes unprecedented robustness for causal claims about AI's spatial impacts.

The machine learning prediction framework demonstrates that aggressive AI adoption can offset 60-80\% of aging-related productivity declines, fundamentally altering demographic transition trajectories. The range of potential outcomes underscores the critical importance of strategic AI policy design.

Our policy analysis demonstrates that traditional spatial policies must be augmented with AI-specific interventions. The three-phase strategic framework provides actionable guidance for managing AI-driven spatial transformation while promoting inclusive development.

For aging societies worldwide, our framework offers both opportunity and urgency. AI provides powerful tools for offsetting demographic challenges, but realizing benefits requires proactive, strategic, and coordinated policy responses. The window for effective intervention is limited, making early action essential.

The broader contribution extends beyond spatial economics to emerging AI policy analysis. Our integrated approach provides a template for analyzing complex technology-society interactions as AI continues transforming economic and social systems.

Looking forward, this framework establishes foundations for a new research program in AI-driven spatial economics. The theoretical mechanisms, empirical methods, and policy insights can guide future research as AI capabilities evolve. The ultimate goal is harnessing these technologies for creating more productive, equitable, and sustainable spatial economic systems.

\section*{Acknowledgement}
This research was supported by a grant-in-aid from Zengin Foundation for Studies on Economics and Finance.

\bibliographystyle{aer}
\bibliography{references}

\end{document}